\documentclass[
 nopreprint
]{jasatex}

\usepackage{color}   
\usepackage{graphicx}
\usepackage{dcolumn}
\usepackage{amsmath,amsfonts}
\usepackage{bm}

\newcommand{\beq}[1]{  \begin{equation} \label{#1} }  
\newcommand{\eeq}{     \end{equation}} 	
\newcommand{\bal}[1]{  \begin{align} \label{#1} }  
\newcommand{\rf}[1]{(\ref{#1})}
\def\bd#1{\mbox{\boldmath$\displaystyle\mathbf{#1}$} }
 
\def\dd{\operatorname{d}} 

\def\grad{\operatorname{grad}} 
\def\div{\operatorname{div}} 
 
\def\axt{\operatorname{axt}} %
\def\curl{\operatorname{curl}}  

\newtheorem{lem}{Lemma}

\def\rev#1{#1}	   
\begin{document} 
\title[Fax\'en relations in solids]{Fax\'en relations in solids - a generalized approach to particle motion in elasticity and viscoelasticity}

\author{Andrew N. Norris}
  \email{norris@rutgers.edu}
\affiliation{Mechanical and Aerospace Engineering, Rutgers University, Piscataway NJ 08854}

\date{\today}

\begin{abstract}
 
A  movable inclusion in an elastic material oscillates as a rigid body  with six degrees of freedom. Displacement/rotation  and force/moment  tensors which  express the motion of the inclusion in terms of the displacement and force at arbitrary exterior points are introduced.  Using reciprocity arguments two general identities are derived relating these tensors.  Applications of the identities to spherical particles provide several new results, including simple expressions for the force and moment on the particle due to plane wave excitation.

\end{abstract}

\maketitle

\section{Introduction}

\rev{
Fax\'en relations  are named after Hilding Fax\'en who derived  several  identities for  calculating hydrodynamic forces and torques on particles in low Reynolds number flows, e.g.  Ref.~ \onlinecite{Faxen27}. 
As an example of a Fax\'en relation, or law, the force and torque on a rigid sphere of radius $a$  moving with  velocity ${\bd v}_0$ and  spinning with angular velocity ${\bd \omega}$ in an unbounded  fluid of viscosity $\mu$ 
and  velocity field ${\bd v}({\bd r})$ in the absence of the sphere 
are [Eqs.  (3-2.46) and (3-2.47) of Ref.~ \onlinecite{Happel91}]
\begin{subequations}
\bal{307}
{\bd F} &= 6\pi \mu a \big[ {\bd v}({\bd 0}) - {\bd v}_0\big]
 + \pi \mu a^3 \nabla^2 {\bd v}({\bd 0}),
\\
{\bd T} &= 4 \pi \mu a^3  \big[ \curl {\bd v}({\bd 0}) - 2 {\bd \omega}\big].
\end{align}
\end{subequations}
The reader will note that the identities have  as a special case the classical Stokes drag law, but  they  include additional   effects caused by spatially variable flow fields. 
These and other  Fax\'en relations for non-spherical particles  are based upon general integral identities relating the force and torque on the particle to the external flow field \cite{Happel91,Kim91,Pozrikidis97}. 
Although Fax\'en relations are commonly used in hydrodynamics and microfluidics,  they  seem to be essentially unknown outside that subject area.  
 For instance, I am aware of only one mention\citep{Phan94} of a Fax\'en type relation in elasticity and that one was is in regards to elastostatics.   
}

\rev{
  The objective of this paper is to develop similar ideas in the context of elastodynamics and in the process demonstrate their utility and wide application.   
  Using dynamic reciprocity, a set of  relations are first derived   
  between the velocity or force of a  particle in a solid  matrix and the displacement or force at a distant point in the solid.   These equations include but go far beyond the notion of particle impedance, which relates the force on a particle to its velocity.  Numerous applications of the  general relations are 
  obtained by considering spherical particles.   
  Fax\'en-like  relations are derived  for the force and moment on a spherical particle caused by plane wave incidence.  Like their hydrodynamic counterparts, the elastodynamic Fax\'en   relations   are simple in form.   
   }
 
The analysis here is the second in a series of papers developing  a simplified algebra for calculating the radiation and  scattering from inclusions in elastic and viscoelastic materials.  \rev{In the  previous paper \citep{norris2006b}  the forced motion of a spherical particle in an elastic matrix was considered.  Although this is a classical problem, originally solved   by Oestreicher \cite{Oestreicher51}, it turns out that the  dynamic impedance of the inclusion can be represented in a simplified manner.  This was achieved \cite{norris2006b} through several 
lumped mass impedances for a spherical inclusion, in terms of which impedance or its inverse, admittance, has a simple form. }
The purpose of the present paper is to develop these ideas further, focusing on the interaction between the inclusion and remote points in the matrix.   


The plan of the paper is as follows.  The dynamic  properties of an inclusion are defined in Section  \ref{sec1a}, as well as  some important quantities that are used throughout the paper: the displacement/rotation tensors $\bd U$ and $\bd W$, the force/moment tensors $\bd \Phi$ and $\bd \Psi$, and the impedances ${\bd Z}$ and $\widetilde{\bd Z}$.  
In Section \ref{sec2} we prove the symmetry of these impedances, and derive two fundamental  relations 
between the displacement/rotation and force/moment tensors.  \rev{The remainder of the paper focuses on the special case of spherical inclusions.  The fundamental quantities for the spherical particle are presented  in Section \ref{sec3} in a concise format using lumped mass  impedances.  Section \ref{sec4} is the longest in the paper, as it contains numerous applications,  discussion of limiting cases, and the new elastodynamic Fax\'en relations that are analogous to the classic hydrodynamic identities.  The many results and their import are summarized in Section \ref{sec5}. }

Regarding notation, the time harmonic factor $e^{-i \omega t}$ is  omitted but understood.  Boldface quantities are either vectors or second order tensors. Vectors are usually denoted by lower case, and tensors are  capitalized, with the exceptions $\bd F$ and $\bd M$ which indicate force and moment vectors, respectively. 
The axial tensor  $\axt ({\bd a})$ of the vector ${\bd a}$ is a skew symmetric tensor defined by $\axt ({\bd a}) {\bd b} = {\bd a}\wedge {\bd b} $.

\section{Inclusions and rigid body motion}\label{sec1a}

An inclusion in a solid matrix is  defined to be the  surface $\partial V_p$ of a finite volume $V_p$  within which there could be a particle, or there could be some complicated ``black box" with its own internal dynamics.  The key feature of the inclusion is that its boundary $\partial V_p$ undergoes rigid body motion.  In this sense  the boundary is a rigid interface between the particle, whatever that may be, and the solid matrix.  The term inclusion rather than particle is used throughout  in order to remind us of this distinction.  

\subsection{Tensor functions ${\bd U}$, ${\bd W}$, ${\bd \Phi}$ and ${\bd \Psi}$}

\begin{figure}[htbp]
				\begin{center}	
				    \includegraphics[width=2.in , height=2.in 					]{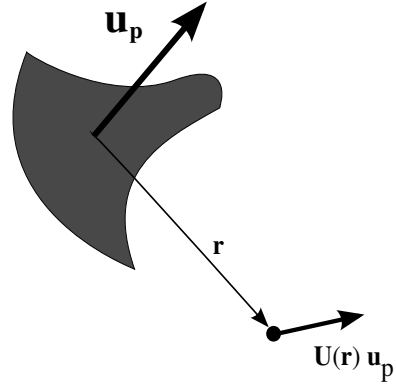} 
	\caption{The inclusion undergoes time harmonic linear displacement ${\bd u}_p$, resulting in displacement ${\bd u} = {\bd U} ({\bd r})  {\bd u}_p$ at position $\bd r$.}
		\label{f1} \end{center}  
	\end{figure}

Rigid body motion has six degrees of freedom, which we characterize by two vector quantities: ${\bd u}_p$ and ${\bd \theta}_p$.   ${\bd u}_p$ is the rigid body displacement of the inclusion center of mass.  ${\bd \theta}_p$ describes the rotation of the inclusion about the center of mass.  The  most general displacement possible for the inclusion is  
\beq{-41}
{\bd u}_P = {\bd u}_p+ {\bd \theta}_p \wedge {\bd r}, \quad \forall {\bd r}\in \partial V_p, 
\eeq
where ${\bd r}$ is the position relative to the center of mass.    We will use the vector ${\bd u}_P$ to denote the total rigid body displacement, with ${\bd u}_p$ reserved for the linear part \rev{(the entire development in this paper applies only to \emph{linear} as distinct from nonlinear motion, so that the term linear is synonymous with rectilinear).}
Note that ${\bd u}_p$ has dimensions of length while ${\bd \theta}_p$  is dimensionless. For the sake of simplicity  it is useful to consider  the linear  and rotational motions separately.    
Figure \ref{f1} shows the inclusion   oscillating back and forth with linear displacement ${\bd u}_p$.  In the absence of other sources of vibrational energy, the inclusion motion induces motion at every point ${\bd r}$ in the  exterior region 
$V = \mathbb{R}^3/ V_p$ according to ${\bd u} = {\bd U} ({\bd r})  {\bd u}_p$, as depicted in Fig. \ref{f1}.  Here ${\bd U} $ is a second order tensor defined everywhere in the matrix.   In the same way  the  particle displacement at ${\bd r}\in V$ caused by  a pure rotation of the inclusion may be defined by a second order tensor $\bd W$.  In short, the tensors $\bd U$ and $\bd W$  relate the rigid body displacement of the inclusion
to the displacement ${\bd u}({\bd r})$ in the exterior solid medium $V$ according to 
\beq{380}
{\bd u}({\bd r}) = {\bd U} ({\bd r})  {\bd u}_p + {\bd W} ({\bd r})  {\bd \theta}_p.  
\eeq
We next define two dual  tensor functions associated with force and moment, respectively.  

\begin{figure}[htbp]
				\begin{center}	
				    \includegraphics[width=2.in , height=2.in 					]{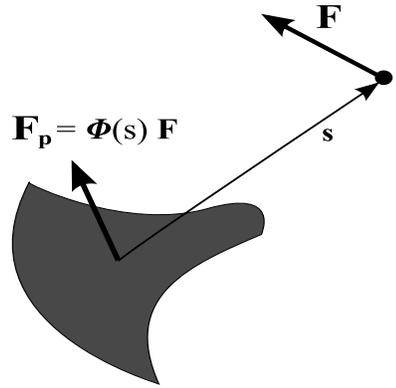} 
	\caption{The time harmonic point force $\bd F$ is applied at $s$, resulting in the force  ${\bd F}_p = {\bd \Phi} ({\bd s})  {\bd F} $ on the inclusion.}
		\label{f2} \end{center}  
	\end{figure}

Consider the situation in which a point force of magnitude times direction equal to ${\bf F}$ acts at ${\bf s}\in V$, inducing motion of the inclusion, Figure \ref{f2}. 
The tensors ${\bd \Phi}$ and ${\bd \Psi}$ define the net force ${\bf F}_p$ and couple ${\bf M}_p$ on the inclusion  caused by the point force according to 
\beq{381}
{\bd F}_p  = {\bd \Phi} ({\bd s})   {\bd F} ,
\qquad   
{\bd M}_p  = {\bd \Psi} ({\bd s})   {\bd F} .
\eeq

\subsubsection{External impedances}

We  introduce two  impedance tensors: ${\bd Z}$ and  $\widetilde{\bd Z}$, called external  impedances because they depend upon the exterior properties of the solid matrix.  
  
\begin{figure}[htbp]
				\begin{center}	
				    \includegraphics[width=2.in , height=2.in 					]{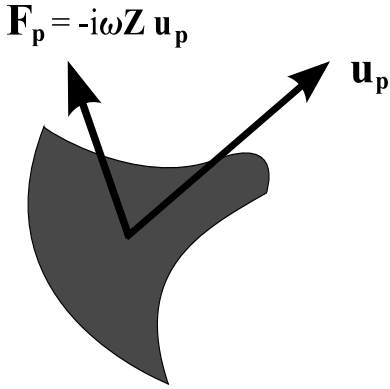} 
	\caption{The inclusion undergoes time harmonic linear displacement ${\bd u}_p$, resulting in the net force ${\bd F}_p$ acting on the inclusion. Conversely, if the force ${\bd F}_p$
	is applied then the displacement is ${\bd u}_p$.}
		\label{f3} \end{center}  
	\end{figure}

The  impedance ${\bd Z}$ relates  the force acting on   the inclusion to the inclusion linear velocity, see Fig. \ref{f3} which is similar to the situation in Fig. \ref{f1}.  It is assumed that either force or velocity  is controlled and the other is the dependent variable, and that there is no other excitation from sources in $V$.  Thus, let ${\bd u}_p$ be  the prescribed inclusion linear  displacement, then the  force ${\bd F}_p$ acting  on the inclusion is
\beq{501}
{\bd F}_p = -i\omega {\bd Z}{\bd u}_p . 
\eeq
${\bd F}_p$ can be thought of as the resultant of the reactive forces from the solid matrix acting on the inclusion, with zero net moment. 
The inverse ${\bd Z}^{-1}$ exists since we may consider the impedance as defined by an imposed force, resulting in the inclusion displacement $ {\bd u}_p
= (-i\omega {\bd Z})^{-1}{\bd F}_p$.   We will prove that ${\bd Z}$ is symmetric (Lemma \ref{lem1}). 

The  moment tensor $\widetilde{\bd Z}$ relates  the moment of the force  on   the inclusion with  the inclusion angular  velocity, $-i\omega \widetilde{\bd Z}{\bd \theta}_p$, 
\beq{5011}
{\bd M}_p = -i\omega \widetilde{\bd Z}{\bd \theta}_p . 
\eeq
$\widetilde{\bd Z}$ is also assumed to be invertible, and  will  be shown to be symmetric (Lemma \ref{lem1}).

\subsubsection{Internal  impedances}

The inertial properties of the inclusion  are defined by two impedance matrices ${\bd Z}_{P}$ 
and $\widetilde{\bd Z}_{P}$ associated with linear and rotational motion, respectively.   We call these internal impedances since they depend entirely on the inclusion and are independent of the exterior region. 

${\bd Z}_{P}$ is a mass-like impedance.  For a normal solid particle it is 
defined by the  mass $m$ as ${\bd Z}_{P} = i\omega m \, {\bd I}$. 
We will generally denote ${\bd Z}_{P}$ as a tensor to include the possibility of  internal structure, although it may be assumed on general principles that the impedance is symmetric, ${\bd Z}_{P} ={\bd Z}_{P}^t$. 

$\widetilde{\bd Z}_{P}$ is the moment of inertia tensor, and is also symmetric
$\widetilde{\bd Z}_{P}= \widetilde{\bd Z}_{P}^t$.  It has  dimensions of a moment of inertia, i.e. mass $\times$ (length)$^2$.

\subsubsection{Summary of main results}

\rev{
The  first principal  result is  a pair of relations  (i) between
 the displacement and force tensors and (ii) 
 between the rotation and moment tensors 
\begin{subequations}
\bal{00}
 {\bd Z}_{P}^{-1} {\bd \Phi} ({\bd r}) &=   ({\bd Z}+{\bd Z}_{P})^{-1} 
   {\bd U}^t ({\bd r}) , 
\\
 \widetilde{\bd Z}_{P}^{-1}{\bd \Psi} ({\bd s}) &=    (\widetilde{\bd Z}+\widetilde{\bd Z}_{P})^{-1} 
   {\bd W}^t ({\bd s}) .  
\label{-425}
\end{align} 
\end{subequations}  }
Thus, ${\bd \Phi}  = {\bd U}^t$ and ${\bd \Psi}  = {\bd W}^t$ if the inclusion is immovable (infinite impedance), however, the relations \rf{00} and \rf{-425} are obviously far more general.   
An immediate  corollary  is that the motion of the inclusion  caused by the remote force at $\bd s$ is 
\begin{subequations}
\bal{001}
{\bd u}_p &= (i\omega )^{-1} ({\bd Z}+{\bd Z}_{P})^{-1} 
   {\bd U}^t ({\bd s}) {\bd F}, 
\\
   {\bd \theta}_p &= (i\omega )^{-1} (\widetilde{\bd Z}+\widetilde{\bd Z}_{P})^{-1} 
   {\bd W}^t ({\bd s}) {\bd F}   .
\end{align}
\end{subequations}
Equations \rf{00} and \rf{-425} are proved in the next Section (Lemma \ref{lem2}). 

\rev{
The second set of principal results concern 
applications to a spherical inclusion in an isotropic matrix.   The displacement and rotations tensors 
 ${\bd U}$ and  ${\bd W}$ have  particularly simple forms when expressed in terms of some lumped parameter impedances introduced in   Ref.~ \onlinecite{norris2006b}.  
Combined with Eqs. \rf{00} and \rf{-425} these lead to a series of useful identities for the force, moment, displacement and rotation of the sphere under different excitation.  For instance, the total force and moment on the sphere caused by a time harmonic   longitudinal or transverse plane wave  is 
\beq{0305}
{\bd F}_p = \Lambda {\bd u}_0,
\qquad
{\bd M}_p = \Gamma {\bd u}_0 \wedge {\bd n},
\eeq
where ${\bd u}_0$ is the wave displacement at the center when the sphere is not present, ${\bd n}$ is the propagation direction,  and the scalars $\Lambda$, $\Gamma$, depend upon the wave frequency,  particle radius,  and other material parameters according to Eqs. \rf{+1}, \rf{+19} and \rf{+20}. 
Equations \rf{0305} could be called Fax\'en relations for solids, by analogy with the use of the term in viscous fluid dynamics.  }

\section{Reciprocity based identities}\label{sec2}

Several  identities are derived in this Section: (i) the symmetry of the external  impedance matrices $\bd Z$ and $\widetilde{\bd Z}$,  (ii)  the relation \rf{00} between  the force and displacement tensors, ${\bd \Phi}$ and ${\bd U}$, and (iii) Eq. \rf{-425} relating the moment and rotation tensors ${\bd \Psi}$ and ${\bd W}$.  The common theme is the use of the dynamic reciprocity.
 
Consider two distinct fields in $V$, labeled $j=1,2$, each with displacement ${\bd u}^{(j)}$, stress ${\bd \sigma}^{(j)}$ and applied body force density per unit volume ${\bd f}^{(j)}$ all in dynamic equilibrium, 
\beq{31}
\div {\bd \sigma}^{(j)} + \rho\omega^2 {\bd u}^{(j)}
+ {\bd f}^{(j)} = 0,
\quad \text{in }\, V. 
\eeq 
The   reciprocity identity (Betti's theorem) follow from standard arguments \cite{Achenbach04}, 
\bal{-32}
 &\int_{\partial V_p}\dd s\,  {\bd u}^{(1)}\cdot {\bd \tau}^{(2)} 
 -\int_V\dd v\,  {\bd u}^{(1)}\cdot {\bd f}^{(2)}  
 \nonumber \\
 & \qquad =
 \int_{\partial V_p}\dd s\,  {\bd u}^{(2)}\cdot {\bd \tau}^{(1)}
 -  \int_V\dd v\,  {\bd u}^{(2)}\cdot {\bd f}^{(1)}   , 
\end{align}
where $\bd \tau$ is the traction vector. 
The surface integrals in Eq.  \rf{-32} involve only quantities in the matrix.  We assume that the following conditions hold at the interface $\partial V_p$: (i) continuity of traction, and (ii) the exterior displacement $\bd u$ is related to the inclusion displacement ${\bd u}_P$ by 
\beq{-66}
{\bd u} - {\bd u}_P= {\bd A}{\bd \tau}, \qquad {\bd r} \in \partial V_p,
\eeq
where ${\bd A}= {\bd A}^t$ is a material parameter.   This spring-like interface condition allows for the possibility of, for instance,  tangential slip, which we will include in the example of the spherical inclusion later.  For the moment we leave ${\bd A}$ as arbitrary. 

The rigid body motion of the inclusion for each of the two distinct solutions in Eq. \rf{-32} is assumed to be a linear displacement ${\bd u}_p^{(j)}$ and a twist ${\bd \theta}_p^{(j)}$, such the total displacement is 
${\bd u}_P^{(j)} = {\bd u}_p^{(j)}+ {\bd \theta}_p^{(j)} \wedge {\bd r}$, see Eq. \rf{-41}.    
Substituting for ${\bd u}^{(j)}$ in the surface integrals then gives
\bal{32}
 &{\bd u}_p^{(1)}\cdot\int_{\partial V_p}\dd s\,   {\bd \tau}^{(2)} 
 + {\bd \theta}_p^{(1)}\cdot\int_{\partial V_p}\dd s\, {\bd r}\wedge  {\bd \tau}^{(2)} 
 \nonumber \\
 & -\int_V\dd v\,  {\bd u}^{(1)}\cdot {\bd f}^{(2)}  
  = -  \int_V\dd v\,  {\bd u}^{(2)}\cdot {\bd f}^{(1)}  
  \nonumber \\
  &
 + {\bd u}_p^{(2)}\cdot  \int_{\partial V_p}\dd s\,  {\bd \tau}^{(1)}  
 + {\bd \theta}_p^{(2)}\cdot\int_{\partial V_p}\dd s\, {\bd r}\wedge  {\bd \tau}^{(1)} 
  .  
\end{align}
Note that the interfacial tensor ${\bd A}$ does not appear in this identity.   We are now ready to derive the fundamental relations,  first considering the impedances. 

\subsection{Symmetry of the external impedances $Z$ and $\widetilde{Z}$ }

Assume that  no force acts in the solid for both solutions, so that 
 ${\bd f}^{(1)} = {\bd f}^{(2)} = 0$.  Then   Eq.  \rf{32} reduces to   
\bal{134}
 &{\bd u}_p^{(1)}\cdot\int_{\partial V_p}\dd s\,   {\bd \tau}^{(2)} 
 + {\bd \theta}_p^{(1)}\cdot\int_{\partial V_p}\dd s\, {\bd r}\wedge  {\bd \tau}^{(2)} 
 \nonumber \\
 & = {\bd u}_p^{(2)}\cdot  \int_{\partial V_p}\dd s\,  {\bd \tau}^{(1)}  
 + {\bd \theta}_p^{(2)}\cdot\int_{\partial V_p}\dd s\, {\bd r}\wedge  {\bd \tau}^{(1)} 
  .  
\end{align} 
The integrals produce the resultant force and moment on the inclusion, which follows from  the  definition of the impedances ${\bd Z}$ and $\widetilde{\bd Z}$ as, 
\begin{subequations}\label{135}
\bal{135a}
 \int_{\partial V_p}\dd s\,  {\bd \tau}^{(j)}  &=- i\omega {\bd Z}{\bd u}_p^{(j)}, 
 \\
  \int_{\partial V_p}\dd s\,  {\bd r} \wedge {\bd \tau}^{(j)}  &=- i\omega \widetilde{\bd Z}{\bd \theta}_p^{(j)},
\end{align}
\end{subequations}
for $j=1,\, 2$. 
The reciprocity relation becomes
\beq{0342}
{\bd u}_p^{(1)} \cdot{\bd Z}{\bd u}_p^{(2)} +{\bd \theta}_p^{(1)} \cdot \widetilde{\bd Z}{\bd \theta}_p^{(2)} = 
{\bd u}_p^{(2)} \cdot{\bd Z}{\bd u}_p^{(1)} +
 {\bd \theta}_p^{(2)} \cdot \widetilde{\bd Z}{\bd \theta}_p^{(1)}.
\eeq

Since ${\bd u}_p^{(1)}$, ${\bd \theta}_p^{(1)}$,   ${\bd u}_p^{(2)}$ and ${\bd \theta}_p^{(2)}$ are arbitrary, we deduce 
\begin{lem}\label{lem1}
The linear and rotational external impedances are symmetric,
\beq{023}
{\bd Z} = {\bd Z}^t,
\qquad
\widetilde{\bd Z} = \widetilde{\bd Z}^t. 
\eeq
\end{lem}

\subsection{Relation between the  force and displacement tensors}

We again take field $1$ as the solution for the inclusion undergoing  arbitrary rigid body   displacement ${\bd u}_P^{(1)}={\bd u}_p^{(1)} +{\bd \theta}_p^{(1)}\wedge {\bd r}$ with  ${\bd f}^{(1)} = 0$. 
Let field $2$ be the  solution for a point force ${\bf F}$ at ${\bd s}$:
\beq{33}
{\bd f}^{(2)}( {\bd x}) =  {\bd F}\, \delta( {\bd x} - {\bd s}),
\quad {\bd s} \in V. 
\eeq
The solution to this, ${\bd u}^{(2)}$, is in fact the 
Green's function in the presence of the movable inclusion.   Our objective is to avoid explicit calculation of the Green's function. 

The displacement   on the inclusion surface  is again a rigid body  displacement, 
${\bd u}^{(2)}_P ={\bd u}^{(2)}_p + {\bd \theta}^{(2)}_p \wedge {\bd r}$, and therefore the reciprocity identity \rf{32} becomes
\bal{34}
 &  {\bd u}_p^{(1)}  \cdot \int_{\partial V_p}\dd s\,  {\bd \tau}^{(2)} 
 +    {\bd \theta}^{(1)}_p \cdot  \int_{\partial V_p}\dd s\,  {\bd r} \wedge {\bd \tau}^{(2)}
 -{\bd F} \cdot   {\bd u}^{(1)} ({\bd s})
 \nonumber \\
 & \quad =
      {\bd u}^{(2)}_p \cdot  \int_{\partial V_p}\dd s\,  {\bd \tau}^{(1)} 
  +    {\bd \theta}^{(2)}_p \cdot  \int_{\partial V_p}\dd s\,  {\bd r} \wedge {\bd \tau}^{(1)} . 
\end{align}
The integrals involving ${\bd \tau}^{(1)} $ again give resultant force and moment according to Eqs. \rf{135} with  $j=1$.   For field $2$, let ${\bd F}_p $ and ${\bd M}_p $ denote the resultants caused by  the point force at $\bd s$, 
\begin{subequations}\label{35}
\bal{35a}
 \int_{\partial V_p}\dd s\,  {\bd \tau}^{(2)}  &= {\bd F}_p ,
 \\
 \int_{\partial V_p}\dd s\,  {\bd r}\wedge {\bd \tau}^{(2)}  &= {\bd M}_p  .
\end{align}
\end{subequations}
 The displacement at $\bf s$ for field $1$ follows from the definition of the tensors $\bd U$
 and $\bd W$ 
as ${\bd u}^{(1)}({\bd s}) = {\bd U}({\bd s})    {\bd u}_p^{(1)}
+{\bd W}({\bd s})    {\bd \theta}_p^{(1)} $, 
see Eq. \rf{380}.

Elimination of these quantities from Eq.  \rf{34} implies 
\bal{39}
 & {\bd F}_p \cdot {\bd u}_p^{(1)}  
 +{\bd M}_p \cdot {\bd \theta}_p^{(1)}  
 -  {\bd F} \cdot   {\bd U} ({\bd s})     {\bd u}_p^{(1)}
 -  {\bd F} \cdot   {\bd W} ({\bd s})     {\bd \theta}_p^{(1)}
\nonumber \\ &
 \qquad =
 - i\omega {\bd u}^{(2)}_p \cdot {\bd Z}     {\bd u}_p^{(1)} 
 - i\omega {\bd \theta}^{(2)}_p \cdot \widetilde{\bd Z}     {\bd \theta}_p^{(1)} . 
\end{align} 
But the rigid body displacement $ {\bd u}_p^{(1)}$ and twist ${\bd \theta}_p^{(1)}$ are arbitrary, and using the symmetry of the impedances, we deduce 
\begin{subequations}\label{40}
\bal{40a}
{\bd F}_p  &=      {\bd U}^t ({\bd s})  {\bd F} 
 -  i\omega {\bd Z}\,   {\bd u}^{(2)}_p      ,
 \\
 {\bd M}_p  &= 
     {\bd W}^t ({\bd s})  {\bd F}
 -  i\omega \widetilde{\bd Z}\,   {\bd \theta}^{(2)}_p  .
\end{align}
\end{subequations} 

A second set of  independent relations follow  from the equilibrium of the inclusion, or Newton's second law applied to a rigid body, 
\begin{subequations}\label{36}
\bal{36a}
{\bd F}_p & = 
i\omega {\bd Z}_{P}   {\bd u}^{(2)}_p ,
\\
 {\bd M}_p & = 
i\omega \widetilde{\bd Z}_{P}   {\bd \theta}^{(2)}_p . 
\end{align}
\end{subequations} 
Eliminating the linear  displacement ${\bd u}^{(2)}_p$  and twist ${\bd \theta}_p^{(2)}$ between Eqs. \rf{40} and \rf{36} gives
\begin{subequations}\label{41}
\bal{41a}
{\bd F}_p  &= {\bd Z}_{P}({\bd Z}+{\bd Z}_{P})^{-1} 
   {\bd U}^t ({\bd s})  {\bd F} ,
   \\
      {\bd M}_p  &= \widetilde{\bd Z}_{P}(\widetilde{\bd Z}+\widetilde{\bd Z}_{P})^{-1} 
   {\bd W}^t ({\bd s})  {\bd F} ,
\end{align}
\end{subequations}
Finally, referring back to the definition of $ {\bd \Phi}$ and $ {\bd \Psi}$ in $\rf{381}$ implies the desired relations:  
\begin{lem}\label{lem2}
The displacement and force   tensors are related by 
\beq{42}
 {\bd \Phi} ({\bd s}) =    {\bd Z}_{P}({\bd Z}+{\bd Z}_{P})^{-1} 
   {\bd U}^t ({\bd s}) . 
\eeq 
The rotation and moment   tensors are related by 
\beq{425}
 {\bd \Psi} ({\bd s}) =    \widetilde{\bd Z}_{P}(\widetilde{\bd Z}+\widetilde{\bd Z}_{P})^{-1} 
   {\bd W}^t ({\bd s}) . 
\eeq 
\end{lem}

We are now ready to examine these quantities for a particular case, the spherical inclusion.  

\section{Spherical inclusion, isotropic matrix }\label{sec3}

\subsection{Definition of the problem}

The inclusion has radius $a$  and is embedded in a uniform isotropic elastic medium of infinite extent with  mass density $\rho$ and Lam\'e moduli $\lambda$ and $\mu$.  
The interface conditions at $r=a$ are: (i) continuity of normal displacement, (ii) satisfaction of a slip condition.  The latter allows for 
relative tangential slip between the inclusion and matrix, and is defined by  
the tangential component of the traction ${\bd \tau}= {\bd \sigma}\hat{\bd r} $  
where 
   ${\bd \sigma}$ is the stress tensor and 
$\hat{\bd r} = r^{-1}{\bd r}$ denotes the unit radial vector.   The tangential component  satisfies 
\beq{67}
{\bd \tau}\cdot \hat{\bd t} = z_I\, (  {\bd v}_P- {\bd v})\cdot \hat{\bd t} ,
\qquad r=a,
\eeq
where $ \hat{\bd t}$ is a  unit tangent vector, $ {\bd v}$ the velocity of the elastic medium adjacent to  the sphere, ${\bd v}_P = -i\omega {\bd u}_P$ is the total velocity of the inclusion at the interface $r=a$, and $z_I$ is an interfacial impedance, introduced in Ref.~ \onlinecite{norris2006b}.  This corresponds to 
${\bd A} = (i\omega z_I)^{-1}({\bd I} - {\bd n}\otimes {\bd n})$ in Eq. \rf{-66}, where $\bd n$ is the interface normal.   The results of Section \ref{sec2} therefore apply for this slip condition. 

In summary, the conditions at the surface of the sphere are 
\beq{091}
\begin{split}
{\bd u} \cdot\hat{\bd r}  &= {\bd u}_P\cdot  \hat{\bd r}  
\\
{\bd \tau} \cdot\hat{\bd t}  &= i\omega z_I\, ( {\bd u} -{\bd u}_P )\cdot \hat{\bd t}
\end{split} 
\, \, 
\Biggr\} 
\qquad r= a . 
\eeq

\subsection{External  impedances}

Symmetry arguments imply that the net force (moment) exerted on the sphere by the surrounding medium  and the resulting linear displacement (axis of rotation) are parallel. 
Hence, the external impedances are isotropic, 
\beq{11}
{\bd Z} = Z {\bd I}, 
\qquad
\widetilde{\bd Z} = \widetilde{Z} {\bd I}\, . 
\eeq
The linear impedance  $Z$ has been considered previously  \cite{norris2006b} while the rotational impedance $\widetilde{Z}$ is new. 
Expressions for both are given next. 
 
 \subsubsection{The linear  impedance}  \label{b1}
The scalar $Z$ can be expressed in a form reminiscent of lumped mass systems \cite{norris2006b} 
\beq{18.1}
\frac3{Z + Z_{M}} = \frac1{Z_L + Z_{M}} +\frac2{Z_S + Z_{M}}\, , 
\eeq
where the  additional impedances are  
\begin{subequations}\label{16}
\begin{align}
Z_{M} &=    i\omega\, \frac43 \pi a^3\rho  ,  
\\
Z_L &=   ( i\omega)^{-1} 4 \pi a(\lambda+2\mu) \, (1-ika),
\label{16b}
\\
Z_T &=   ( i\omega)^{-1}  \, 4 \pi a\mu (1-iha), \label{16c}
\\
\frac1{Z_S} &= \frac1{Z_T} + \frac1{  4 \pi a^2 z_I + (i\omega)^{-1}  \, 8 \pi a\mu } . 
\label{16d}
\end{align}
\end{subequations}
Here $k$ and $h$ are, respectively, the longitudinal and transverse wavenumbers, $k=\omega/c_L$, $h=\omega/c_T$ with $c_L = \sqrt{(\lambda + 2\mu)/\rho }$ and $c_T = \sqrt{\mu/\rho }$.  The impedances in \rf{16}  depend upon and are defined by the matrix properties, except for $Z_S$, which involves the interface viscosity term $z_I$.
Thus, $Z_M$ is the mass-like impedance of a sphere of the matrix material of the same size as the inclusion.    
Note that $Z_S = Z_T$ if the inclusion is perfectly bonded to the matrix ($z_I \rightarrow \infty$).  See Ref.~ \onlinecite{norris2006b} for further discussion of this and other limits.

\subsubsection{The rotational impedance}  \label{b2}

The rotational impedance of a spherical inclusion has not, to our knowledge, been presented in the literature.  A  derivation is given in Appendix \ref{appa}, with the result that  
${\widetilde{Z}}$ is 
\beq{033}
\frac{a^2}{\widetilde{Z}} =  \frac{3}{8\pi a^2 z_I} + 
\frac{\tfrac12 (1 - i ha)}{Z_M + Z_T}.
\eeq
The parameters in this identity were defined previously. 

\subsection{Displacement, rotation,  force and moment tensors}

\subsubsection{Internal impedance}

The  internal impedances ${\bd Z}_{P}$ and $\widetilde{\bd Z}_{P}$ are necessary in order to relate the displacement/rotation  tensors with the force/moment tensors via Lemma \ref{lem2}. 
For the sake of simplicity we restrict consideration in this paper to  internal impedances that are  isotropic: 
\beq{-111}
{\bd Z}_{P} = Z_{P}{\bd I}, 
\qquad
\widetilde{\bd Z}_{P} = \widetilde{Z}_{P}{\bd I}.
\eeq
For instance, a uniformly solid   sphere of mass $m$ has 
\beq{-112}
Z_{P} = i\omega m,
\qquad
\widetilde{Z}_{P} = i\omega \frac25 a^2 m.
\eeq

\subsubsection{Linear motion}

The   displacement  tensor $\bd U$ of Eq. \rf{380} is derived in  Appendix \ref{appb} as
\bal{47}
 &{\bd U} ({\bd r}) =    a
 (Z+Z_M   )  \bigg[ 
 \frac{- 1}{Z_L+Z_M}  \, 
 \nabla\nabla \frac{ e^{ik(r-a)} }{k^2 r}
 \nonumber \\ 
 & \quad
 +  \frac{1 }{Z_S+Z_M} \frac{Z_S}{Z_T}\, 
 \big( \nabla\nabla +h^2 {\bd I}\big) \frac{ e^{ih(r-a)} }{h^2 r} \bigg],
\quad r\ge a .  
\end{align}
The force tensor   $\bd \Phi$ of Eq. \rf{381} follows from Lemma \ref{lem2} and the fact that the impedances satisfy  \rf{11} and \rf{-111}.  Thus, 
\beq{64}
 {\bd \Phi} ({\bd r}) =     \frac{Z_{P}}{Z+ Z_{P}} 
 {\bd U} ({\bd r}).
 \eeq

The displacement and force tensors satisfy ${\bd U}(-{\bd r})  = {\bd U}({\bd r})$, 
${\bd \Phi}(-{\bd r})  = {\bd \Phi}({\bd r})$
and  are symmetric, 
${\bd U} = {\bd U}^t $,
$ {\bd \Phi}=  {\bd \Phi}^t$.
We focus on the properties of ${\bd U}$ since those of $ {\bd \Phi}$  are easily obtained through \rf{64}. 

Equation \rf{47} implies 
\bal{042}
 {\bd U}  = &   
 \frac{Z+Z_M}{Z_L+Z_M}  \big[ \frac{h_1(kr)}{kr h_0(ka)} {\bd I} -\frac{h_2(kr)}{h_0(ka)} \hat{\bd r}\otimes \hat{\bd r}
\big]
\nonumber \\
&  
+  \frac{Z+Z_M}{Z_S+Z_M} \frac{Z_S}{Z_T}
 \big( [\frac{h_0(hr)}{h_0(ha)}-\frac{h_1(hr)}{hr h_0(ha)}] {\bd I} 
 \nonumber \\
&  
+ \frac{h_2(hr)}{h_0(ha)} \hat{\bd r}\otimes \hat{\bd r} \big),  
\end{align}
where
 $h_n$ are spherical Hankel functions of the first kind \cite{Abramowitz74}
 and  $\hat{\bd r} = r^{-1}{\bd r}$ denotes the unit radial vector.   In particular  $h_0(z) = (iz)^{-1}e^{iz}$. 
In  expanded form, 
\begin{widetext}
\bal{471}
 {\bd U} ({\bd r}) =     
 (Z+Z_M ) \, \frac{a}{r} \bigg[ &
 \frac{1}{Z_L+Z_M}  \big[ (\frac{1}{(kr)^2} -\frac{i}{kr}){\bd I} 
 +(1+\frac{3i}{kr}-\frac{3}{(kr)^2} ) \hat{\bd r}\otimes \hat{\bd r}
\big] e^{ik(r-a)}
\nonumber \\
& 
+  \frac{1}{Z_S+Z_M} \frac{Z_S}{Z_T}
 \big[ (1+\frac{i}{hr}-\frac{1}{(hr)^2} )  {\bd I} 
- (1+\frac{3i}{hr}-\frac{3}{(hr)^2} )  \hat{\bd r}\otimes \hat{\bd r} \big]e^{ih(r-a)} \bigg].
\end{align}  
\end{widetext}

\subsubsection{Rotational  motion}

The skew tensor $\bd W$ relating the rotation to the displacement at a distance  follows from Appendix \ref{appa}  as 
\beq{-553}
{\bd W}( {\bd r}) = - a \Omega \frac{h_1(hr)}{h_1(ha)}\, \axt(\hat{\bd r}), 
\quad
\Omega = 1 - \frac{3\widetilde{Z} }{8\pi a^4 z_I}. 
\eeq
Then $\bd \Psi$, which relates the moment on the inclusion to an applied force at a distance, is
\beq{-64}
 {\bd \Psi} ({\bd r}) =     \frac{- \widetilde{Z}_{P}}{\widetilde{Z}+ \widetilde{Z}_{P}} 
 {\bd W} ({\bd r}).
 \eeq
The rotational tensors satisfy are odd functions of their arguments, ${\bd W}(-{\bd r})  = -{\bd W}({\bd r})$, ${\bd \Psi}(-{\bd r})  = -{\bd \Psi}({\bd r})$, 
and  are skew symmetric, 
${\bd W} = -{\bd W}^t $,
$ {\bd \Psi}=  -{\bd \Psi}^t $.

\section{Applications}\label{sec4}

This Section explores implications of the general theory to the particular case of the spherical inclusion. 

\subsection{Force on a particle from  plane wave incidence}

The force on a particle due a remote point load is given directly by the tensor $\Phi ({\bd r})$. 
Taking the source point to infinity the effect of the excitation on the particle is equivalent to an incident plane wave, or a combination of two incident plane waves.   The far-field form of $\Phi ({\bd r})$ follows from Eqs. \rf{64} and \rf{471} as  
\bal{-471}
 {\bd \Phi} ({\bd r}) =&    Z_{P} \bigg(
 \frac{Z+Z_M}{Z + Z_{P}}  \bigg) 
 \frac{a}{r}
   \bigg[    \frac{ e^{ik(r-a)}}{Z_L+Z_M}  \hat{\bd r}\otimes \hat{\bd r}  
   \nonumber \\
   &
 +  \frac{e^{ih(r-a)} }{Z_S+Z_M} \frac{Z_S}{Z_T}
 \big(  {\bd I} -  \hat{\bd r}\otimes \hat{\bd r} \big) \bigg] 
 + \text{O}(r^{-2}).
\end{align}
At the same time, the far-field free space Green's function is (see Eq. \rf{92}), 
\beq{=92}
{\bd G}^{(0)} ({\bd r}) = \frac1{4\pi \mu r} \bigg[ \kappa^{-2} e^{ikr} 
\hat{\bd r}\otimes \hat{\bd r}   +   e^{ihr} \big(  {\bd I} -  \hat{\bd r}\otimes \hat{\bd r} \big)
 \bigg]
 + \text{O}(r^{-2}).
\eeq 

Consider, for instance, a unit point force in the far-field  at $\bd r$ in the direction ${\bd n} = -\hat{\bd r}$ .  This produces a longitudinal plane wave at the origin  of the form ${\bd u} = u_0 {\bd n}e^{ik{\bd n}\cdot {\bd x}}$ where $u_0 = e^{ikr}/(4\pi \mu \kappa^2 r)$.  The force on the spherical particle due to an incident longitudinal plane wave 
 \beq{--}
 {\bd u} ({\bd x}) =e^{ik{\bd n}\cdot {\bd x}}\,  {\bd u}_0,
 \qquad  {\bd u}_0\wedge {\bd n} =0, 
\eeq
is therefore
\beq{+1}
 {\bd F}_p =      (\lambda + 2\mu)     
 \frac{ 4\pi a Z_{P}}{Z + Z_{P}}     \bigg( \frac{Z+Z_M}{Z_L + Z_M}  \bigg) \,e^{-ika} \, {\bd u}_0. 
\eeq

In the same manner, the force on the spherical particle due to an incident transverse plane wave 
 \beq{-4}
 {\bd u} ({\bd x}) =e^{ih{\bd n}\cdot {\bd x}}\,  {\bd u}_0 , \qquad {\bd u}_0\cdot {\bd n}=0, 
\eeq
is
\beq{+19}
 {\bd F}_p =      \mu   \, 
 \frac{4\pi a Z_{P}}{Z + Z_{P}}    \frac{Z_S}{Z_T} \bigg( \frac{Z+Z_M}{Z_S + Z_M}  \bigg)  \, e^{-iha}\,  {\bd u}_0. 
\eeq
The  values of the plane wave induced forces for the rigid immovable particle 
follow from Eqs. \rf{+1} and \rf{+19} in the  limit as $Z_{P} \rightarrow \infty$.   These  values actually coincide in the static limit, as discussed below after we consider the quasistatic limit of $Z$. 

 \rev{Nagem and Davis \cite{Davis06} considered plane wave incidence on an 
elastic sphere in a compressible viscous fluid, with specific results focused on the   rigid immovable limit.}  This is equivalent to an isotropic elastic medium with shear modulus $\mu = -i \omega \rho \nu_0$, where $\nu_0$ is the kinematic viscosity, and with a viscously damped longitudinal wave.    Their expression for the force on the rigid sphere  under acoustic plane wave incidence (Eqs. (30) and (31) of Ref.~ \onlinecite{Davis06}) should  agree  with \rf{+1} in the rigid  limit. 
 
 \subsection{Moment on a particle from   a plane wave }
 
 The far-field form of the moment tensor is, from Eqs. \rf{-553} and \rf{-64}, 
 \beq{-14}
 {\bd \Psi} ( {\bd r}) =  \frac{a^2 \Omega} {1-(iha)^{-1}} \frac{e^{ih(r-a)}}{r}\,  \frac{\widetilde{Z}_{P}}{\widetilde{Z}+ \widetilde{Z}_{P}} \, \axt(\hat{\bd r}) . 
 \eeq
 Based on the discussion for the forcing from plane wave incidence, it is evident that a longitudinal wave produces zero net moment on the spherical particle.   A transverse plane wave, does however, exert a moment.  It may be shown that the plane wave \rf{-4} produces  
\beq{+20}
 {\bd M}_p =         \frac{ 4\pi a^2 \mu } {1-(iha)^{-1}} \,  \frac{\Omega \widetilde{Z}_{P}}{\widetilde{Z}+ \widetilde{Z}_{P}} \, e^{-iha}\,  {\bd u}_0 \wedge{\bd n}. 
\eeq
The rigid and quasistatic limits are discussed below. 

\subsection{Rigid body  displacement due to   a plane wave }

The  particle displacement under plane wave incidence is a combination of a linear displacement and a rigid body rotation.  These follow, respectively,  from the forcing ${\bd F}_p $ of Eqs. \rf{+1} or \rf{+19} and the moment ${\bd M}_p $ of Eq. \rf{+20} as 
\beq{+190}
 {\bd u}_P =    (i\omega  Z_{P} )^{-1} {\bd F}_p 
 + (i\omega  \widetilde{Z}_{P} )^{-1} {\bd M}_p \wedge {\bd r}, \quad r\le a.  
\eeq
Symmetry dictates that the moment tensor ${\bd M}_p$ is zero for longitudinal incidence.

\subsection{Quasistatic limit } 

 \subsubsection{Linear motion: Virtual mass}
\begin{figure}[htbp]
				\begin{center}	
				    \includegraphics[width=2.5in , height=2.5in 					]{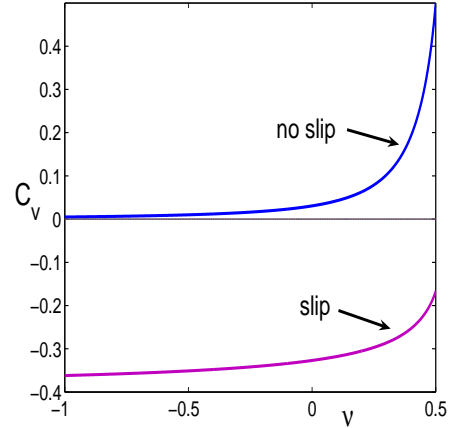} 
	\caption{The virtual mass coefficient $C_v$ of the spherical inclusion  plotted as a function of the Poisson's ratio.  The no-slip and slip curves correspond to $\chi=0$ and $\chi = 1$ in 
	 Eq. \rf{-432}, respectively. }
		\label{f4} \end{center}  
	\end{figure}

The quasistatic limit of vanishingly small frequency  $(\omega \rightarrow 0)$ yields 
\bal{19}
Z =&   \frac{12\pi a \mu  }{i\omega} 
\bigg[ \frac{1 }{2+ \chi +\kappa^{-2} }  
       -      \frac{i ha (2+ \kappa^{-3}) }{(2+ \chi +\kappa^{-2})^2 }
  \nonumber \\
   &  -  \frac{ C_v}{9 } (ha)^2 
+ {\text O}(h^3a^3)\bigg]\, , \quad |ha|, |ka|\ll 1,
\end{align}
where 
\beq{08}
\kappa = \frac{c_L}{c_T} = \sqrt{ \frac{ 2(1-\nu)}{1-2\nu} }, 
\eeq
and $\nu$ is the Poisson's ratio.  The non-dimensional factor 
 $\chi$  is related to the interface impedance $z_I$ in this limit, and is chosen so that it takes on the values zero or unity in the limit that the sphere is either perfectly bonded or  perfectly lubricated, 
\beq{12}
\chi = 
\begin{cases}
0, & \text{no slip}, \quad z_I\rightarrow \infty, \\
1, & \text{slip}, \quad \quad \  z_I = 0.
\end{cases}
\eeq
The parameter $C_v$ is 
\beq{-432}
C_v = 9\frac{(2+ \kappa^{-3})^2 }{(2+ \chi +\kappa^{-2})^3  }
-6 \frac{ (2-\frac54 x+ \kappa^{-4})}{(2+ \chi +\kappa^{-2})^2}    - 1 . 
\eeq
The expansion \rf{19} goes further than in 
 Ref.~ \onlinecite{norris2006b} (Eq. (30)) which did not contain the $C_v$ term.   If the low frequency expansion is of the form 
$Z = Z^{(-1)}( i\omega)^{-1} + Z^{(0)} + Z^{(1)}( i\omega) + \ldots $ then the 
coefficient $Z^{(1)}$ determines the  extra inertia or added mass caused by the linear motion of the infinite matrix. The virtual mass coefficient   is defined as  
$Z^{(1)}/Z_M$, and is therefore $C_v $ of Eq. \rf{-432}.    As shown in  Fig. \ref{f4}, the 
coefficient is positive under no slip conditions for all permissible values of Poisson's ratio.  It approaches the limiting value of $C_v =1/2$ in the limit of incompressibility, $\nu \rightarrow 1/2$, in agreement with the value for viscous fluids \cite{Kendoush05}.  In contrast,  the virtual mass coefficient is always negative when the inclusion is permitted to slip, and is always less than the incompressible limiting value of $C_v =-1/6$.

 \subsubsection{Linear motion: Static displacement}

Based on Eq. \rf{19}  we have  
\bal{89}
 {\bd U} =& 
  \frac{3 a }{2+ \chi +\kappa^{-2} } 
  \bigg[ \frac{1}{r} {\bd I}  + 
  \lim_{h\rightarrow 0}\, \nabla\nabla 
 \frac{1}{h^2 r} \bigg(  \frac{ e^{ih(r-a)} } {1-i ha-\frac13 h^2a^2 }
  \nonumber \\
  & \qquad 
 - \frac{e^{i\kappa^{-1} h(r-a)} } {1-i \kappa^{-1}  ha-\frac13 \kappa^{-2}h^2a^2 } \bigg) \bigg]
 \nonumber \\
  = &
  \frac{3 a }{2+ \chi +\kappa^{-2} } 
  \big[ \frac{1}{r} {\bd I}  + \frac12 (\kappa^{-2} - 1)
 \nabla\nabla 
  \big(  r + \frac{a^2}{3r}
 \big) \big]. 
 \end{align}
 Evaluating the gradients and removing $\kappa$ in favor of $\nu$ gives 
 \bal{90}
 {\bd U} ({\bd r}) =& 
  \frac{3 a }{2r[5-6\nu +2\chi (1-\nu)] } 
  \big[ \big( 3-4\nu +\frac{a^2}{3r^2} \big) {\bd I}  
   \nonumber \\
   &
  + \big( 1-\frac{a^2}{r^2} \big)  \hat{\bd r}\otimes \hat{\bd r}\big],
  \quad \text{static \& immovable}.
 \end{align}

This can be compared with  Walpole's \cite{Walpole05} result for the static perfectly bonded immovable spherical inclusion, Eq. (3.21) of Ref.~ \onlinecite{Walpole05}.  He considered the force tensor ${\bd \Phi}$, which as we have seen is equal to ${\bd U}^t$ in the limit of a fixed and immovable rigid inclusion, i.e. $Z_{P}\rightarrow \infty$.   But ${\bd U}$ is symmetric for the sphere, and therefore Eq. \rf{90} represents both ${\bd U}$ and ${\bd \Phi}$. 
When  $\chi = 0$ this  agrees with Walpole.  The static result for  $\chi = 1$ appears to be new. 

A more precise definition of the fixed inclusion limit is that $Z_{P}/Z \rightarrow \infty$.  At the same time we are taking the static limit, so the simultaneous static and immovable limit is 
\beq{235}
\lim\limits_{\omega \rightarrow 0} \, \frac{a\mu}{\omega Z_{P} } = 0. 
\eeq

 \subsubsection{Rotational motion: Virtual mass}
 
 In the limit of low frequency   the rotational impedance 
 ${\widetilde{Z}}$ of Eq.  \rf{033} for approximates as
\beq{-12}
\widetilde{Z} = 
\begin{cases}
\frac{8\pi a^3 \mu}{i\omega} + i\omega \frac{8}3 \pi a^5 \rho + \text{O}(\omega^2), & \text{no slip},  \\
0, & \text{slip}, 
\end{cases}
\eeq
where slip and no slip correspond to the limits $z_I = 0$ and $z_I \rightarrow \infty$, respectively. 
The term $i\omega \frac{8}3 \pi a^5 \rho = 2a^2 Z_M$ can be identified as the virtual mass due to the rotating solid.  The internal rotational impedance of a solid sphere is 
$\widetilde{Z}_{P} = \frac25 a^2 {Z}_{P}$.  Hence the virtual mass in rotation is five times the mass of solid  matrix material in the volume of the sphere, in agreement with a similar result for Stokes flow \cite{Kendoush05}.
 
\subsubsection{Quasistatic  plane wave force on an immovable particle}

The forcing on the particle from plane wave incidence is the same, whether longitudinal or transverse waves are incident, in the  quasistatic limit for a rigid immovable sphere.  It may be checked that both \rf{+1} and \rf{+19} become 
\beq{-19}
 {\bd F}_p =       
 \frac{12\pi a  \mu}{2+ \chi + \kappa^{-2}}   \,  {\bd u}_0. 
\eeq
This apparently strange result may be reconciled with the physical nature of the limit: the matrix moves by the static displacement $ {\bd u}_0$, while the particle is stationary.  It is therefore sensible that there is a net force on the particle, and that it is in the direction of  $ {\bd u}_0$.  This also explains the identical form of the limit for both types of wave incidence.    The static limit  is one of the unusual features of the the immovable sphere.  For further discussion   see Pao \cite{Pao73}, who  quite properly  questions the  physical validity of the rigid fixed assumption.  Among its failings, as Pao notes, this configuration   does not display Rayleigh scattering behavior at low frequencies. 

\subsubsection{Quasistatic  plane wave moment on a fixed sphere}

The  moment on the  rigid immovable spherical  particle  has a particularly simple form, 
\beq{+21}
 {\bd M}_p =      \frac{-iha}{1-iha}4\pi a^2   \mu  \,\Omega \, e^{-iha}\,  {\bd u}_0 \wedge{\bd n}, 
 \quad \text{immovable}. 
\eeq
This is valid at all frequencies, but as $\omega \rightarrow 0$ it vanishes, unlike the force on the particle in the same limit.

\subsection{Small inclusion limit}

\subsubsection{Linear motion}
As $a \rightarrow 0$ we have $Z$, $Z_L$, $Z_S$, $Z_T$ = O$(a)$, while 
$Z_M$, $Z_{P}$ = o$(a)$. It may be easily verified that $\bd U$ and  $\bd \Phi$  reduce in this limit to 
 \beq{91}
 {\bd U} ({\bd r}) = i\omega Z\, {\bd G}^{(0)} ({\bd r}), 
 \qquad
  {\bd \Phi} ({\bd r}) = i\omega Z_{P}\, {\bd G}^{(0)} ({\bd r}), 
 \eeq
where   ${\bd G}^{(0)}$ is the free space Green's tensor 
\beq{92}
{\bd G}^{(0)} ({\bd r}) = \frac1{4\pi \mu} \bigg[
\frac{ e^{ihr} }{  r} {\bd I} +
 \nabla\nabla \bigg(  \frac1{h^2 r}\big( e^{ihr}  -  e^{ikr} \big) \bigg)  
 \bigg].
\eeq 
In hindsight, the form of ${\bd \Phi} $ is obvious based on the dynamic  equilibrium of the inclusion : $i\omega Z_{P}{\bd u}_p =  {\bd F}_p$ with ${\bd u}_p = {\bd G}^{(0)} {\bd F}$,  ${\bd F}_p = {\bd \Phi}{\bd F}$,  and $\bd U$ then follows from \rf{64}.  

\subsubsection{Rotational motion}
The rotational quantities ${\bd W}$ and ${\bd \Psi}$, on the other hand, become negligible in the small inclusion limit.  This follows from the scaling ${\bd W} = $O$(a)$ in Eq. \rf{-553}. 

\subsection{Surface displacement and traction} 

\subsubsection{Linear motion}
The displacement tensor $\bd U$, which is defined in the exterior region, reduces to the following on the interface:
\beq{043}
 {\bd U}  (a \hat{\bd r}) =      {\bd I} 
 + \big(\frac{Z_S-Z_T}{Z_T}\big)  \big(\frac{Z+Z_M}{Z_S+Z_M}\big)  
 \big(  {\bd I} - \hat{\bd r}\otimes \hat{\bd r} \big) . 
\eeq
This becomes the identity ${\bd I}$ under no-slip conditions, since then $Z_S-Z_T= 0$. 
Alternatively, the interface conditions \rf{091} can be written 
\beq{0911}
{\bd u} = {\bd u}_p +(i\omega z_I)^{-1} (\cdot\hat{\bd t} \otimes \hat{\bd t})
{\bd \tau}  ,
\qquad r= a   .
\eeq
The final term on the RHS vanishes as $z_I\rightarrow \infty$, which is the no-slip limit.

Substituting  ${\bd u} =  {\bd U}  (a \hat{\bd r}){\bd u}_p $ on $r=a$, in \rf{0911} provides an explicit expression for  the interfacial shear traction in terms of the linear displacement ${\bd u}_p $,
\beq{881}
{\bd \tau} \cdot\hat{\bd t}  = i\omega z_I\, 
 \big(\frac{Z_S-Z_T}{Z_T}\big)  \big(\frac{Z+Z_M}{Z_S+Z_M}\big)\, 
  {\bd u}_p \cdot\hat{\bd t} .
 \eeq
The shear traction vanishes under pure-slip conditions $(z_I=0)$, and for a bonded interface it becomes 
 \beq{8815}
{\bd \tau} \cdot\hat{\bd t}  = -i\omega {\bd u}_p \cdot\hat{\bd t} \, \frac{ Z_T}{4\pi a^2} \, 
  \big(\frac{Z+Z_M}{Z_T+Z_M}\big)  
  . 
 \eeq
 
 \subsubsection{Rotational motion}
 
The displacement on $r=a$ is, 
\beq{-556}
{\bd W}( a\hat {\bd r}) = - a \Omega \, \axt(\hat{\bd r}), 
\eeq
where $\Omega$, given in  \rf{-553}, reduces to unity under no-slip conditions.  Conversely,
$\Omega =0$  for pure slip, indicating that the solid does not move even as the inclusion rotates.  
 
In this case the traction is pure shear, and 
 \beq{-051}
  {\bd \tau}  =  -i\omega  \frac{3 \widetilde{Z} }{8\pi a^3}\, {\bd \theta}_p \wedge \hat{\bd r}, 
  \qquad r=a.
\eeq
This is non-zero except under pure slip conditions, when $\widetilde{Z} \rightarrow 0$, and there is no rotational  interaction between the inclusion and the matrix.

\subsection{Acoustic limit}

\subsubsection{General formulation}
Finally, we discuss how the general elastodynamic formulation reduces when the matrix  is an acoustic fluid. In this limit shear effects are ignorable and the medium is characterized  by density $\rho$ and  bulk modulus $K=\rho c^2$, where $c$ is the acoustic wave speed.  Taking the displacement $\bd u$ and pressure $p$ as field variables, the momentum balance and constitutive law are, respectively, 
\beq{601}
\omega^2 \rho {\bd u} =   \nabla p,
\qquad
p =  - K \nabla \cdot {\bd u}. 
\eeq 
The acoustic wavenumber is $k=\omega /c$. 

We introduce two vector functions ${\bd q}({\bd r})$ and ${\bd \phi}({\bd s})$ that are analogous to the tensors ${\bd U}$  and ${\bd \Phi}$.  If the inclusion is moved back and forth with the displacement   ${\bd u}_p$ then the condition on the inclusion surface is that the normal velocity is continuous, 
\beq{602}
{\bd u} \cdot {\bd n} = {\bd u}_p \cdot {\bd n} \quad \text{on } \partial V_p. 
\eeq
The pressure at a point ${\bd r}$ away from inclusion is defined by ${\bd q}$ as
\beq{603}
p ({\bd r}) =  {\bd q}({\bd r}) \cdot {\bd u}_p \quad {\bd r} \text{in }   V. 
\eeq
Conversely, consider a voluminal source at ${\bd s}$: 
\beq{604}
\nabla^2 p + k^2 p = f \delta( {\bd x} - {\bd s}).  
\eeq
The force on the inclusion is 
\beq{605}
{\bd F}_p = - 
 \int\limits_{\partial V_p}\dd s\,  p \, {\bd n}    
  \equiv f\,  {\bd \phi}  ( {\bd s})  ,
 \eeq
which defines the vector function ${\bd \phi}$. 

The connection  between ${\bd q}$ and ${\bd \phi}$ is given by 
\begin{lem}\label{lem2a}
The acoustic displacement and force   vectors are related by 
\beq{606}
 {\bd \phi} ({\bd s}) =   (\rho\omega^2)^{-1}\,  {\bd Z}_{P}({\bd Z}+{\bd Z}_{P})^{-1} 
   {\bd q} ({\bd s}) . 
\eeq 
\end{lem}
This may be derived by  application of  reciprocity to the acoustic (Helmholtz) equation, in a manner similar to how we derived Lemma \ref{lem2}.

\subsubsection{Spherical inclusion}

Finally, we consider the example of the spherical inclusion.  The vector ${\bd q} $ follows from the acoustic limit of the elastic result in \rf{47}, 
\beq{617}
 {\bd q} ({\bd r}) =    aK
   \big( 
 \frac{Z+Z_M  }{Z_A+Z_M}  \big)\, 
 \nabla\frac{ e^{ik(r-a)} }{ r},
\quad r\ge a ,
\eeq
where $Z_A$, analogous to the longitudinal impedance $Z_L$ in elasticity, is 
\beq{618}
Z_A=   ( i\omega)^{-1} 4 \pi aK \, (1-ika), 
\eeq
$Z_M$ is as before, and the sphere impedance $Z$ is now given by \rf{18.1} with $Z_S = 0$, which implies 
\beq{18.2}
\frac1{Z} = \frac2{Z_{M}}+ \frac3{Z_A} .
\eeq

\section{Conclusion}\label{sec5}

Starting from the  notion of an inclusion with the  six degrees of freedom of a rigid body, we  introduced displacement/rotation  and force/moment  tensors relating the motion of the inclusion to the displacement and force at arbitrary exterior points.  These can be considered as generalized Green's functions appropriate to the constrained nature of the inclusion.  The  general  relations \rf{00} and \rf{-425} between the displacement/rotation  and force/moment  tensors are one of the main contributions of the paper.     These identities are extremely useful  in providing a means by which one can consider the dynamic properties of  particles embedded in a solid matrix.  

Useful results have been obtained for  the simplest but important  configuration of a  uniform spherical particle.    The linear  and rotational impedances, $Z$ and $\widetilde{Z}$, are given in Sections \ref{b1} and \ref{b2}, respectively, the latter for the first time.   Explicit expressions are given in Eqs. \rf{64} - \rf{-64} for the displacement tensors ${\bd U}$ and ${\bd W}$  and for the force tensors ${\bd \Phi }$ and ${\bd \Psi }$. Perhaps the most practical new results are  Eqs. \rf{+1}, \rf{+19} and  \rf{+20} which provide simple formulae for 
the force and moment on a particle under plane wave incidence.     The associated displacement of the particle is given by Eq.  \rf{+190}.  These concise expressions resemble Fax\'en relations that are frequently used in microhydrodynamics.

Equation \rf{19} extends the quasistatic expansion of Norris \cite{norris2006b} to include the virtual mass coefficient, which can be negative if slip occurs.   The quasistatic form of ${\bd U}$, Eq. \rf{90}, which relates the displacement of the inclusion to  particle displacement in the matrix,   generalizes a recent formula of 
Walpole \cite{Walpole05}.  The quasistatic limit of the plane wave force on a sphere, Eq. \rf{-19}, is reminiscent of Stokes drag law, but is proportional to the displacement vector of  the incident plane wave.  It also includes the possibility of slip relative to the matrix.  However, the  moment on a rigid sphere from plane wave incidence is proportional to the incident particle velocity, Eq.  
\rf{+21}, and vanishes in the limit of zero frequency.   Other 
limiting cases considered include the small inclusion limit, and the purely acoustic limit.  

Taken together the results of paper offer a consistent means for analyzing wave-particle interaction in elasticity and viscoelasticity.  
Future applications will look at replacing the solid  spherical particle   with more complicated, and more interesting, internal structure.  This amounts to considering more general forms of the internal impedances.  The results developed here can also be used to develop simplified methods for  scattering from particles.  These issues will be examined in separate papers.

\appendix  

 \section{Rotation  of a sphere}\label{appa}
 
 The sphere $r\le a$ undergoes oscillatory rotation ${\bd u}_P = {\bd \theta}_p\wedge {\bd r}$.  Let ${\bd e}$ be the axis of rotation, so that ${\bd \theta}_p = \theta_p {\bd e}$, and consider the possible solution ${\bd u} = \nabla \wedge  {\bd e} f$ in the matrix $r>a$. 
 This satisfies the equations of motion
 \beq{-45}
 {\bd u} + k^{-2} \nabla\nabla\cdot  {\bd u} -h^{-2} \nabla\wedge \nabla\wedge {\bd u}
 =0
 , 
 \eeq
provided that  $f$ is a solution of the reduced Helmholtz equation 
$ \nabla^2 f  + h^{2} f=0$. 
The function $f$ must depend only on $r$ in order to match the prescribed rotation on $r=a$. Hence,  
 \beq{-44}
 {\bd u} =  
 \beta h_1(hr)\, {\bd e} \wedge \hat{\bd r}, \quad r>a, 
 \eeq
where $\hat{\bd r} =  {\bd r}/r$.  
The traction in an isotropic solid  is \cite{Oestreicher51},
\beq{05}
  {\bd \tau}  = \hat{\bd r} \lambda \div  {\bd u} + \frac{\mu}{r} \grad {\bd r}\cdot {\bd u}
 + \mu \big(  \frac{\partial }{\partial r} - \frac1{r}\big) {\bd u},
\eeq
from which we obtain 
\beq{051}
  {\bd \tau}  = \beta \mu h \big[ h_0(ha) - \frac{3}{ha} h_1(ha) \big] {\bd e} \wedge \hat{\bd r}.
\eeq
The boundary conditions \rf{091} therefore reduce to a single equation for the parameter $\beta$ of Eq. \rf{-44}, 
\beq{052}
 \beta \mu h \big[ h_0(ha) - \frac{3}{ha} h_1(ha) \big] = 
 i\omega z_I \big[ \beta h_1(ha)  - a \theta_p \big]. 
\eeq

The moment ${\bd M}_p = \int  \dd s {\bd r} \wedge {\bd \tau}$ is obtained  from the identity
\beq{44-}
\int\limits_{r=a} \dd s \, {\bd r} \wedge ( {\bd e}\wedge \hat{\bd r})
= \frac83 \pi a^3 {\bd e}, 
\eeq
and the impedance then follows from the definition \rf{5011} as 
$\widetilde{\bd Z} = \widetilde{Z}{\bd I}$ where $\widetilde{Z}$ is given by \rf{033}.

\section{Displacement of a sphere}\label{appb}

The solution to the radiation boundary value problem of the spherical inclusion undergoing linear displacement with boundary conditions defined by \rf{091} has been solved by Oestreicher \cite{Oestreicher51} for the case of no slip and more recently, by Norris \cite{norris2006b}, with slip included.  We follow the latter with some slight changes in notation.   The solution is based on the following  representation  for the  elastic field outside the sphere, $r\ge a$, 
\beq{02}
{\bd u} = - C_1 \, {\bd u}_p\cdot \nabla\nabla h_0(kr)
 +  C_2 \, {\bd u}_p \cdot \big( \nabla\nabla - {\bd I}\nabla^2\big) h_0(hr), 
\eeq
where $r = |{\bd r}|$ is the spherical radius  and $h_0$ is the spherical Hankel function of the first kind, 
 $h_0(z) = (iz)^{-1}e^{iz}$. 
 
 In the notation of Ref.~ \onlinecite{norris2006b}, $C_1 = -A_1/k^2$ and 
$C_2 = 3B_1/h^2$.  Also, with reference to Norris \cite{norris2006b}, the inclusion displacement is of magnitude $u_0$: ${\bd u}_p = u_0 \, \hat{\bd x}$. 
 Equations (15)
and (19a) of  Ref.~ \onlinecite{norris2006b} combined with Eq. \rf{11}, give  (noting that $Z_m$ of 
Ref.~ \onlinecite{norris2006b} is now $Z_M$) 
\begin{subequations}\label{-101}
\begin{align}
h_2(ka) A_1 &= \big(1+\frac{Z}{Z_M}\big) u_0, 
\\
6\frac{h_1 (ha)}{ha}B_1 & = \frac{h_1 (ka)}{ka}A_1 - \frac{Z}{Z_M} u_0 . 
\end{align}
\end{subequations}
Using 
 $h_1(z) = -h_0'(z)$ and  $h_2 = -h_0 + 3z^{-1}h_1$ implies the identities 
 \beq{-2}
  \frac{h_1(ka)}{ka h_0(ka)} = -\frac{Z_L}{3Z_M},
    \quad
 \frac{h_2(ka)}{h_0(ka)} = - \,\frac{Z_L+Z_M}{Z_M} . 
\eeq
Similar identities  for arguments $ha$ instead of $ka$ have $Z_T$ instead of $Z_L$. 
Combining these results implies
\begin{subequations}\label{101}
\begin{align}
C_1 &= \big( \frac{Z+Z_M}{Z_L+Z_M}\big)\, \frac{ 1}{k^2h_0(ka)}, 
\\
C_2 & = \big(\frac{Z+Z_M}{Z_S+Z_M}\big)\frac{Z_S}{Z_T} \, \frac{ 1}{h^2 h_0(ha)} . 
\end{align}
\end{subequations}
 The expression  \rf{47} for $\bf U$ then follows. 
 

\end{document}